\documentclass{article}
\usepackage[version=3]{mhchem}
\usepackage[left=1.5cm, right=1.5cm, top=1.785cm, bottom=2.0cm]{geometry}
\usepackage{sectsty}
\usepackage{graphicx} 
\usepackage{lastpage}
\usepackage{float}
\usepackage{setspace}
\usepackage{authblk}

\newcommand{\vect}[1]{\boldsymbol{#1}}

\begin{document}

%
%
%
%
%
%


\title{Controlled propulsion and separation of helical particles at the nanoscale}

\author[1,*]{Maria Michiko Alcanzare}
\author[1]{Vaibhav Thakore}
\author[2]{Santtu T. T. Ollila}
\author[3]{Mikko Karttunen}
\author[1,4]{Tapio Ala-Nissila}
\affil[1]{COMP CoE at the Department of Applied Physics, Aalto University School of Science, P.O. Box 11000, 
FIN-00076 Aalto, Espoo, Finland}
\affil[2]{Varian Medical Systems Finland, Paciuksenkatu 21, 00270 Helsinki, Finland}
\affil[3]{Department of Mathematics and Computer Science \& Institute for Complex Molecular Systems, Eindhoven University of Technology, P.O.Box 513, MetaForum 5600 MB Netherlands}
\affil[4]{Department of Physics, Box 1843, Brown University, Providence, Rhode Island 02912-1843, USA}
\affil[*]{maria.alcanzare@aalto.fi}





\maketitle
\begin{abstract}
Controlling the motion of nano and microscale objects in a fluid environment is a key factor in designing optimized tiny machines that perform mechanical tasks such as transport of drugs or genetic material in cells, fluid mixing to accelerate chemical reactions, and cargo transport in microfluidic chips. Directed motion is made possible by the coupled translational and rotational motion of asymmetric particles. A current challenge in achieving directed and controlled motion at the nanoscale lies in overcoming random Brownian motion due to thermal fluctuations in the fluid. We use a hybrid lattice-Boltzmann Molecular Dynamics method with full hydrodynamic interactions and thermal fluctuations to demonstrate that controlled propulsion of individual nanohelices in an aqueous environment is possible. We optimize the propulsion velocity and the efficiency of externally driven nanohelices. We quantify the importance of the thermal effects on the directed motion by calculating the P\'eclet number for various shapes, number of turns and pitch lengths of the helices. Consistent with the experimental microscale separation of chiral objects, our results indicate that in the presence of thermal fluctuations at P\'eclet numbers $>10$, chiral particles follow the direction of propagation according to its handedness and the direction of the applied torque making separation of chiral particles possible at the nanoscale. Our results provide criteria for the design and control of helical machines at the nanoscale.

\end{abstract}



Most natural organisms and molecules exhibit chiral structures that significantly influence their chemical and physical behavior. For example, most proteins are left-handed while sugars are right-handed. Drug molecules which target proteins, therefore, have different effects depending on their chirality. Thalidomide is perhaps the most infamous example of a drug whose devastating health effects have been ascribed to the presence of both left and right handed enantiomers \cite{franks2004thalidomide}. Motion and kinetics of chiral objects are largely determined by their spatial anisotropy since it leads to a coupling between the translational and rotational (TR) degrees of freedom. In biological systems, TR coupling is utilized by microorganisms that propel themselves in fluids by rotating their flagella and generating a chirality. 

Artificially manufactured chiral particles with coupled TR motion provide many opportunities for technological applications
\cite{Walker,ghosh2009controlled,tottori2012magnetic,sumigawa2015substrate,schamel2013chiral,medina2015cellular}.
Recently, Clemens \textit{et al.} have experimentally demonstrated separation of racemic mixtures by exposing left and right-handed chiral molecules to a rotating electric field. The dipole moments of the molecules in the racemic mixture align with the rotating electric field leading to their rotational motion. This propels the chiral molecules via the TR coupling and gives rise to enantiomer separation depending on their handedness \cite{clemens2015molecular}. 
Controlled and directed motion of chiral particles has important applications in targeted delivery of genetic material and drugs, too, because such particles can be tailored to facilitate their efficient uptake by various cells or tissues \cite{singh2009nanoparticle}. Tottori \textit{et al.} demonstrated controlled motion and cargo transport of magnetic helical particles \cite{tottori2012magnetic}. Application of time-dependent magnetic fields on magnetic screw-like particles induces a torque that rotates the particle and propels it in the fluid. Manipulation of the direction of the rotating magnetic fields makes it possible to load microspheres as cargo into the holder of a helical particle, transport the cargo and then release it at a preferred site \cite{tottori2012magnetic}.  Medina-S{\'a}nchez \textit{et al.} have even demonstrated the possibility of assisted sperm delivery on immotile live sperm using externally driven microhelices \cite{medina2015cellular}. Walker \textit{et al.}, on the other hand, focused on optimizing the propulsion velocity of helices with a payload by varying the helical length \cite{Walker}. In the experiments above, the propellers are typically $200-300$ nm wide and $1-2$ $\mu$m long \cite{Walker,ghosh2009controlled,tottori2012magnetic,sumigawa2015substrate}. 

On the theoretical side, exact analytic results for the drag coefficients of certain high-symmetry shapes such as spheres and ellipsoids can be obtained in the Stokes limit. For more complex shapes such as chiral particles, the boundary integral formulation of the Stokes equation can be used \cite{keaveny2013optimization,ghosh2009controlled,zhang2009artificial}. The typical rotational Reynolds number in experiments of driven magnetic helices at the micron scale and at the nanoscale are about $10^{-8}$ and $10^{-9}$ \cite{schamel2014nanopropellers}. Hence zero Reynolds number approximation is justified. The resistive force theory (RFT) and the slender body theory (SBT) are the usual approaches used to study the propulsive motion of rigid helices and flagellar propulsion. The RFT works under the assumption that the segments of the helical body do not generate fluid disturbances and consequently these segments do not influence the motion of the other segments of the helical body. The hydrodynamic forces on an infinitesimal segment are proportional to the local velocity of the body \cite{johnson1979flagellar} by drag coefficients. The calculations for the propulsive velocity of rigid helices are carried out by summing the drag and thrust contributions of the rotating helical particles \cite{de2011simplified,raz2007swimming}. The main weakness of the RFT lies in the neglect of the long-range hydrodynamic interactions. The SBT incorporates these hydrodynamic interactions by considering the fluid response due to the moving segments of the helical body by assuming a local point force represented by Stokeslets that are uniformly distributed along the center line of the helical body. To account for the no-slip boundary condition, corresponding doublets are also uniformly distributed along the center line \cite{johnson1979flagellar,rodenborn2013propulsion,Hancock96,lighthill1976flagellar,wada2009hydrodynamics}. The regularized SBT assumes that the force is not a Dirac delta at a single point but uses a shape function that modulates the force to be spatially distributed in a small blob around a point \cite{cortez2005method}. In terms of the propulsion of microorganisms by a helical flagellum, the numerical results of the SBT and the regularized SBT are similar \cite{rodenborn2013propulsion}. 

Recently, artificial helical particles at the nanoscale with radii less than $100$ nm have become available \cite{schamel2014nanopropellers,sone2002semiconductor,gibbs2013plasmonic,eslami2014chiral}. Controlled propulsion has been experimentally demonstrated in a biological gel and in a high viscosity glycerol-water mixture (25 cP) but at low viscosities such as pure water, Brownian motion was greater than the propulsion of the nanohelix such that no directed motion was observed \cite{schamel2014nanopropellers}. The challenge in controlling nanohelices is overcoming thermal effects which can significantly change the direction of motion and hinder the propulsion. Their coupled TR motion in aqueous solutions have not been characterized yet. Open questions include how the chiral geometry and the fluid properties affect the efficiency and the propulsion speed in cargo delivery. Here, we address these issues by employing the recently developed fluctuating Lattice-Boltzmann - Molecular Dynamics (LBMD) method \cite{Ollila2011,Ollilasiam,mackay2013coupling,mackay2013hydrodynamic,citelammps}. The method incorporates full Navier-Stokes hydrodynamics with consistent thermal fluctuations and a coupling of the fluid to extended MD particles of arbitrary shapes. Since its development, the fluctuating LBMD has been extensively benchmarked for colloids and polymers \cite{Ollila2011,Ollilasiam,ollila2013hydrodynamic,mackay2014modeling}.
We consider externally driven helices with radii of $30$ nm, optimize the shape parameters and identify the conditions for controlled and directed motion, and chiral separation of helices at the nanoscale in general (Typical values of the Reynolds numbers in the simulations are shown in the Fig. S1). Our results provide clear criteria for the physical parameters such as the P\'eclet number and fluid viscosity essential to studying applications of driven helical particles in nanoscale systems.

\begin{figure}[h]
\centering
\includegraphics[width=0.35\textwidth]{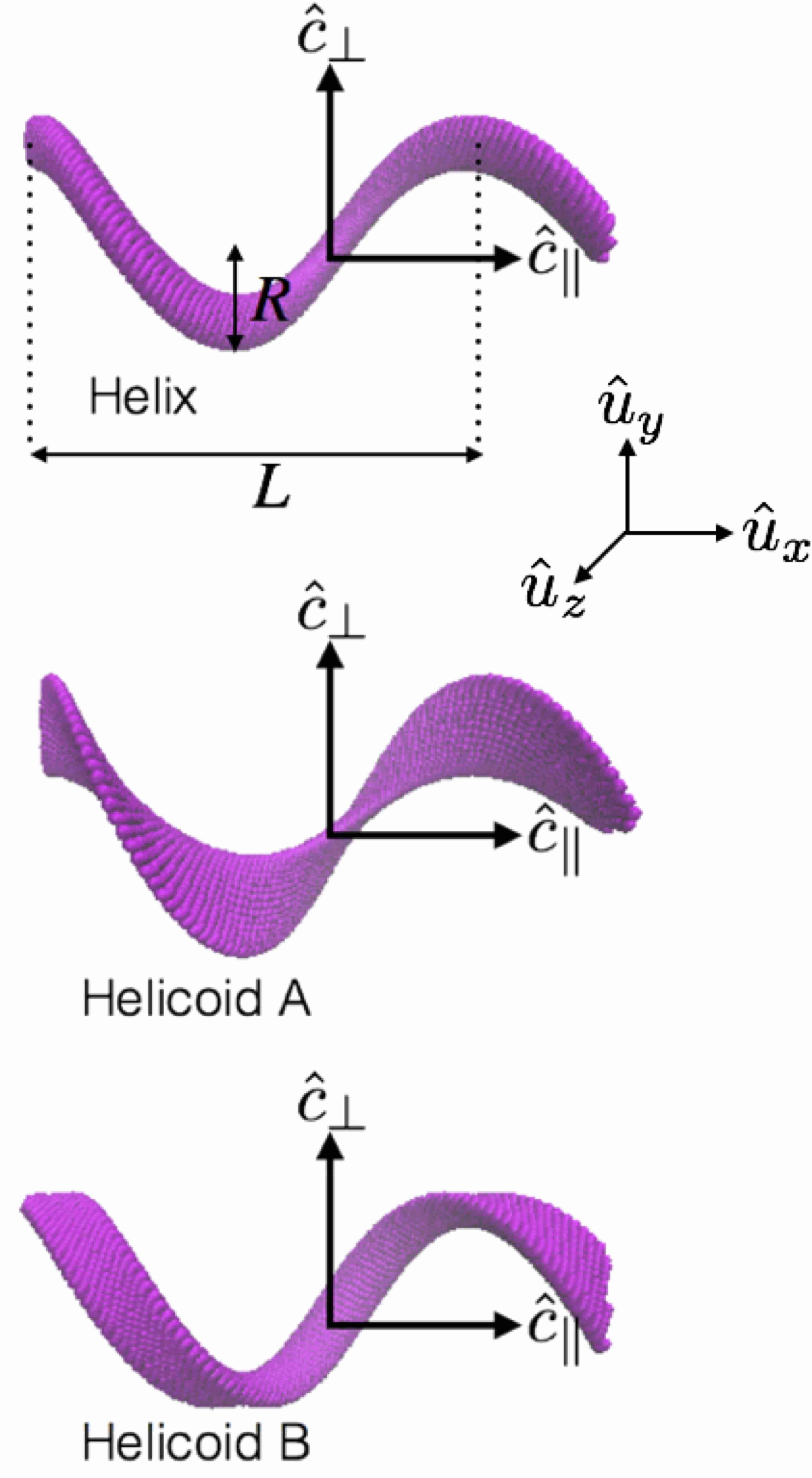}
\caption{\textbf{Geometry of the helical particles.} 
Shapes of different helices with a helical radius of $R$, helical turns $N=1.25$ and pitch length $L$. The unit vectors
$\hat{u}_x$, $\hat{u}_y$ and $\hat{u}_z$ represent coordinates in the laboratory frame, while $\hat{c}_\perp$ and $\hat{c}_\perp$ represent the coordinates in the body frame of the helical particles.
}
\label{directions}
\end{figure}
\section*{Results and Discussion}
\subsection*{Shape optimization of the helical particles}

The fundamental quantity to be considered in the driven motion of helical particles is the propulsion velocity $v$ induced by the rotational motion from the application of a constant external torque. The angular velocity of the helix for a constant torque depends on viscous drag coefficients which are presented in Fig. S2. We first present results for optimizing the geometric shape parameters for the helical particles shown in Fig. \ref{directions} such that $v$ is maximized by using Method 1 as described in the Methods section. During optimization the hydrodynamic interactions are fully taken into account, but thermal noise is neglected.

\begin{figure*}[h]
\centering
\includegraphics[width=0.95\textwidth]{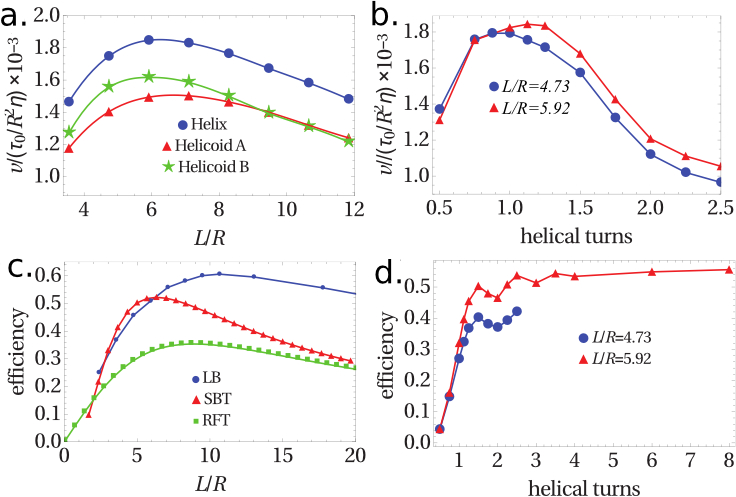}
\caption{\textbf{Optimization of the helical particles.} 
a. Propulsion velocities as a function of $L/R$ of the helix, helicoid A, and helicoid B for $N=1.25$ with a constant torque of $\tau_0=1.0\times10^{-18}$ kg m$^2$/s$^2$. $L/R\approx6$ results in maximum propulsion velocities for the helix, and helicoids A and B.
b. Propulsion velocities as a function of the number of helical turns of a helix with $L/R=4.73$ and $L/R=5.92$. The applied constant torque is $1.0\times10^{-18}$ kg m$^2/$s$^2$. c. Efficiency of helices with $N=1.25$ from the simulations using the hybrid lattice Boltzmann Molecular Dynamics method (circles), from the slender body theory (triangles) and from the resistive force theory (squares) as a function of $L/R$.
d. Efficiency $\epsilon\equiv v/(\omega R)$ of helices of pitch length and radius ratio of 4.73 and 5.92.
}
\label{fig2}
\end{figure*}

The velocity $v$ depends on several parameters. The drag force experienced by a helical particle as it propels in the fluid depends on the shape of its cross-section. We consider here three different shapes as shown in the Methods, 
where the helix has a circular cross-section and helicoids A and B have elliptic cross-sections. The helicoid A (B) has its long axis parallel (perpendicular) to the radial direction of the long axis (see diagram in Fig. \ref{directions}). The volumes of the helical particles are kept constant here and the elliptic cross-sections of helicoids A and B have a linear eccentricity of $\sqrt{3}/2$. Other relevant parameters are the number of helical turns $N$ and the pitch length (defined as the length $L$ for one full turn of a helix). We keep the helical radius fixed at $R=30$ nm and thus optimize the ratio $L/R$. In the shape optimization a fluid density and viscosity of water at room temperature are used as $\rho=998$ kg/m$^{3}$ and $\eta=1.0$ mPa s, respectively. 
 

Comparing the different helical particles in Fig. \ref{directions} we find that for all the pitch lengths considered and for the same applied external torque, the angular velocity of the helix is the highest of them all (cf. Fig. S3) because it displaces the least fluid during rotation. Also, the helix yields the greatest propulsion velocity when the pitch length is varied, as shown in Fig. \ref{fig2}.a because it has the smallest circular area of cross-section resulting in the lowest drag along the long axis. The maximum propulsion velocity for the helix is obtained around $L/R \approx 6$ (Fig. \ref{fig2}.a) for
$N=1.25$. Keaveny and co-workers optimized the shape of propellers with an attached spherical cargo for maximum propulsion. Using the boundary integral formulation $L/R\approx4$ ($L/R\approx6$) they observe maximum propulsion for a helical filament radius and helical radius ratio of 0.2 (0.4) \cite{keaveny2013optimization}. As shown in Fig. S4, due to the strong spatial asymmetry of the helices there is significant wobbling for $N < 1.125$. In order to achieve maximum propulsion, both wobbling and the drag force must be minimized, and thus $N$ must be larger than unity. In Fig. \ref{fig2}.b we show the propulsion velocity for two values of $L/R$ in the region of the velocity maximum as a function of $N$, showing that the optimal number of turns is indeed $N \approx 1.25$. This is consistent with the experimental result of Walker \textit{et al.} \cite{Walker}, where maximum propulsion is achieved for about one full helical turn for micron sized particles.

\subsubsection*{Propulsion efficiency}

In addition to maximizing the propulsion velocity, it is also useful to characterize the efficiency of the helix.
The dynamics at the low Reynolds number of a particle is described by the generalized resistance tensor
\begin{equation}
\left(\begin{matrix}
\vect{F} \\ \vect{\tau}
\end{matrix}\right)
=
\left( \begin{matrix}  \vect\xi^\text{TT} & \vect\xi^\text{TR} \\ \vect\xi^\text{RT} & \vect\xi^\text{RR} \\
\end{matrix} \right)
\left(\begin{matrix}
\vect v \\ \vect\omega
\end{matrix}\right),
\end{equation}
where $\vect F, \vect \tau, \vect v, \vect \omega$ and $\vect\xi^{\alpha\beta}$ are the net force, torque, translational velocity, angular velocity and the friction matrix of the particle, respectively \cite{purcell1977life}. Particles with spherical symmetry have zero non-diagonal terms in the friction matrix and diagonal terms correspond to the Stokes translational and rotational drag coefficients. For a helix, when a torque is applied along its long axis, the rotational motion drives it forward along this same axis through the non-zero coupling term $\vect\xi^{TR}$ or $\vect\xi^{RT}$. To characterize the TR coupling or the amount of propulsion for a given angular velocity, 
we define the efficiency $\epsilon$ to be the constant of proportionality of the propulsion velocity and the angular velocity, $\epsilon \equiv v/(\omega R)$, where $R$ is the helical radius. 

Within the resistive force theory (RFT), there exists an approximate analytic result for the (relative) efficiency given by

\begin{equation}
\epsilon' = \frac{\sin 2\theta}{3-\cos 2\theta}\label{RFT-efficiency}
\end{equation}
where $\theta$ is the pitch angle that is related to the pitch length and the helical radius by $\tan\theta = 2\pi R/L$ \cite{de2011simplified,raz2007swimming}. In Fig. \ref{fig2}.c we compare the numerically obtained results for the efficiency of a helix ($N=1.25$) to that in Eq. (2) and the efficiency calculated using the slender-body theory (for the efficiencies of different shapes see Fig. S5). In our data, the maximum efficiency is achieved at $L/R\approx11.7$, which is about $1.3$ and $1.8$ times larger than that predicted by the RFT and SBT, respectively. These results clearly justify the use of the quantitatively accurate LBMD method to calculate the efficiency. 

In Fig. \ref{fig2}.d we show numerical results for the efficiency as a function of the number of turns $N$ for two values of $L/R$ close to the optimum in Fig. \ref{fig2}.a. In contrast to Eq. (2), where there is no $N$ dependence, with full hydrodynamics we find nontrivial dependence of $\epsilon$ on $N$. The stability of the rotational motion and propulsion, and consequently the efficiency, increases with the number of turns from $N=0.5$ to $N=1.5$, where there is a local maximum. Beyond $N=1.25$, the efficiency exhibits damped oscillations before saturating to a constant value, where both $\omega$ and $v$ asymptotically
decrease as $N^{-1}$ (see Fig. S5).

\subsection*{Influence of thermal fluctuations}
Thermal fluctuations affect the motion of the helices by changing the orientation of the helix and interfere with the translational and rotational motion. In this section we characterize the strength of the thermal fluctuations with respect to the rotational and translational motion. 

For the rotational motion, the rotational P\'eclet number characterizes the strength of the driven rotation and the thermal effects, $\text{Pe}_R=\omega/D_R$, where $D_R$ is the rotational diffusion coefficient along the long axis which is the ratio of the thermal energy and the viscous rotational drag coefficient $D_R=k_B T/\kappa_\text{rot}$. At steady state, the external torque is proportional to the angular coefficient with a constant of proportionality of $\kappa_\text{rot}$, and therefore $\text{Pe}_R=\tau/(k_B T)$. In all of the simulations of the driven helix, $\tau>48 k_B T$ hence the strength of the thermal fluctuations is weaker compared to the induced rotation by the external torque. The effects of the thermal fluctuations on the rotational motion were observed to be negligible. In ESI Fig. S6, we show the orientations of the perpendicular axis in the presence of thermal fluctuations for the applied external torques that are $\geq100\times10^{-21}$ kg m$^2$/s$^2$.

\subsubsection*{Directed motion and chiral separation}

For the directed motion of the helical particles, the translational P\'{e}clet number  ($\text{Pe}_T\equiv vL/D_T$, where $D_T$ is the Brownian tracer diffusion coefficient of the particle) quantifies the strength of the propulsive versus random diffusive motion and is thus a relevant measure of thermal fluctuations. It can be written as
\begin{equation}
{\rm Pe}(\omega) = \frac{ 6 \pi \eta R_H}{k_B T} \epsilon \omega R L,
\label{Pecletnum}
\end{equation}
where $R_H$ is the effective size (hydrodynamic radius) of the particle. 

As can been seen from Eq. (\ref{Pecletnum}), the effect of temperature is typically negligible for micron size particles due to the large Pe for relevant velocities, and has not been considered to date.
However, at the nanoscale Pe is not necessarily large and thermal fluctuations need to be fully accounted for. 

\begin{figure}[h!]
\centering
\includegraphics[height=7cm]{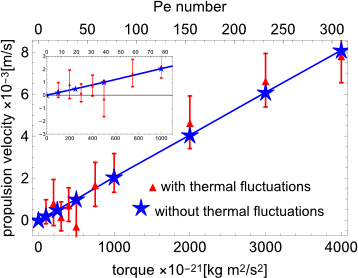}
\caption{\textbf{Propulsion velocities at various torque strengths.} 
Average propulsion velocities for varying torque strength of a helix with $N=1.25$ helical turns and $L/R=5.92$ in simulations without thermal fluctuations, and with thermal fluctuations. Inset shows details of
the data for the small Pe number regime.}
\label{method1-method2-propulsion-velocities-thermalized}
\end{figure}

In the ESI (Fig. S9) we show how $D_T$ depends on $N$. As expected, Pe grows linearly
with the applied torque (Fig. S10.a). To obtain directed motion, $N$ must be greater than unity to minimize wobbling. In Fig. \ref{method1-method2-propulsion-velocities-thermalized} we present data for the propulsion velocity of an optimal nanohelix ($N=1.25$, $R=30$ nm and $L/R = 5.92$) as a function of the external torque with and without thermal fluctuations. The velocities are averaged over the ratio of the displacements and a time interval of $\delta t=4.4\times10^{-6}$ s for 80 independent trajectories. From the data we can conclude that directed motion becomes very well established for P\'eclet numbers greater than approximately $50$.


\begin{figure*}[h]
\centering
\includegraphics[width=0.95\textwidth]{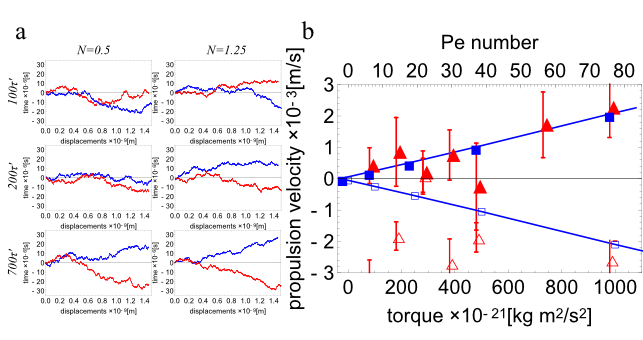} 
\caption{\textbf{Chiral separation.} 
a. Typical displacement trajectories vs. time
of a positive (blue) and a negative (red) helix with a pitch length and helical radius ratio of $L/R=5.92$ that is driven by various torque strengths in the presence of thermal fluctuations, with a constant 
torque indicated in the figure ($\tau'=10^{-21}$ kg m$^2/$s$^2$).
b. Measured average propulsion velocities of helices with $N=1.25$ helical turns and $L/R=5.92$ with thermal fluctuations (triangles) and without thermal fluctuations (squares). Filled (empty) markers have positive (negative) chirality. 
}
\label{displacement-thermalized-pn-helix-rev1p25}
\end{figure*}

Separation of the magnetic helical colloids with a radius $R\approx 0.3$ $\mu$m and number of turns $N=5$ has been experimentally demonstrated at high ratios ($\sim10^3$) of torque strengths and thermal energies \cite{schamel2013chiral}. We present results for the nanoscale helices driven at constant torque ($R=30$ nm and $L/R=5.92$) with positive $(+)$ and negative $(-)$ helicities with thermal fluctuations at $300$ K. 

In the absence of the external torque, purely diffusive motion is observed. At constant external torque in the presence of thermal fluctuations, chiral particles with $(+)$ and $(-)$ helicities move in the two opposite directions parallel to their long axis. 
In Fig. \ref{displacement-thermalized-pn-helix-rev1p25}.a, we show typical displacement trajectories of 
the $(+)$ and $(-)$ helices with $0.5$ and $1.25$ helical turns for various torque strengths. At $\tau_\text{max}=1.0\times10^{-19}$ kg m$^2/$s$^2$, the angular velocity obtained is not sufficient to separate the chiral particles with $N=0.5$ and $N=1.25$. At a higher torque $\tau_\text{max}=2.0\times10^{-19}$ kg m$^2/$s$^2$, 
the 
longer helical particles ($N=1.25$) get separated while chiral separation for the 
short particles is still not possible. In general, we conclude that P\'eclet numbers below about $10$ are 
definitely not sufficient to separate chiral particles. This result is consistent with the 
experimentally demonstrated chiral separation of microscale bacteria in sheared flows \cite{fu2009separation}.

\section*{Summary and Conclusions} 
We have shown here that helical particles with a circular cross-section characterized by a minimum surface area result in optimal propulsion in terms of maximising the propulsion velocity. 
A circular cross-section of the flagella, often naturally encountered in many organisms, is thus optimized for propulsion. Controlled propulsion in the presence of thermal fluctuations requires stability in the rotational motion (\textit{i.e.}, minimum wobbling) which increases with the number of turns $N$. However, increase in the length adds to the hydrodynamic drag. Therefore, maximum propulsion is achieved by balancing the competing requirements for maintaining the stability of the rotational motion and minimizing viscous drag. Thus, the optimal number of turns, $N=1.25$, required for obtaining maximum propulsion velocities in the high P\'eclet number limit reported here, is consistent with the previously published experimental results for microscale helices \cite{Walker}. 
At the nanoscale where thermal effects are significant, we have shown that for well-defined
directed motion and chiral separation of magnetic particles, the P\'eclet number should definitely be larger than $10$, in good agreement with the experiments on the chiral separation of microscale objects reported earlier \cite{fu2009separation}. 
\section*{Methods}

\subsection*{Hybrid Lattice-Boltzmann - Molecular Dynamics method}

The multiscale hybrid Lattice-Boltzmann - Molecular Dynamics (LBMD) method implemented in 
LAMMPS is used in simulating the propulsion of the helical nanoparticles in the fluid \cite{Ollila2011,Ollilasiam,mackay2013coupling,mackay2013hydrodynamic,citelammps}. 
The method incorporates full Navier-Stokes hydrodynamics with consistent coupling of the MD particles
to thermal fluctuations in the fluid. The method has been extensively benchmarked and tested for colloids and polymers \cite{Ollila2011,Ollilasiam,ollila2013hydrodynamic,mackay2014modeling}.

\begin{figure}[htpb]
\centering
\includegraphics[width=0.45\textwidth]{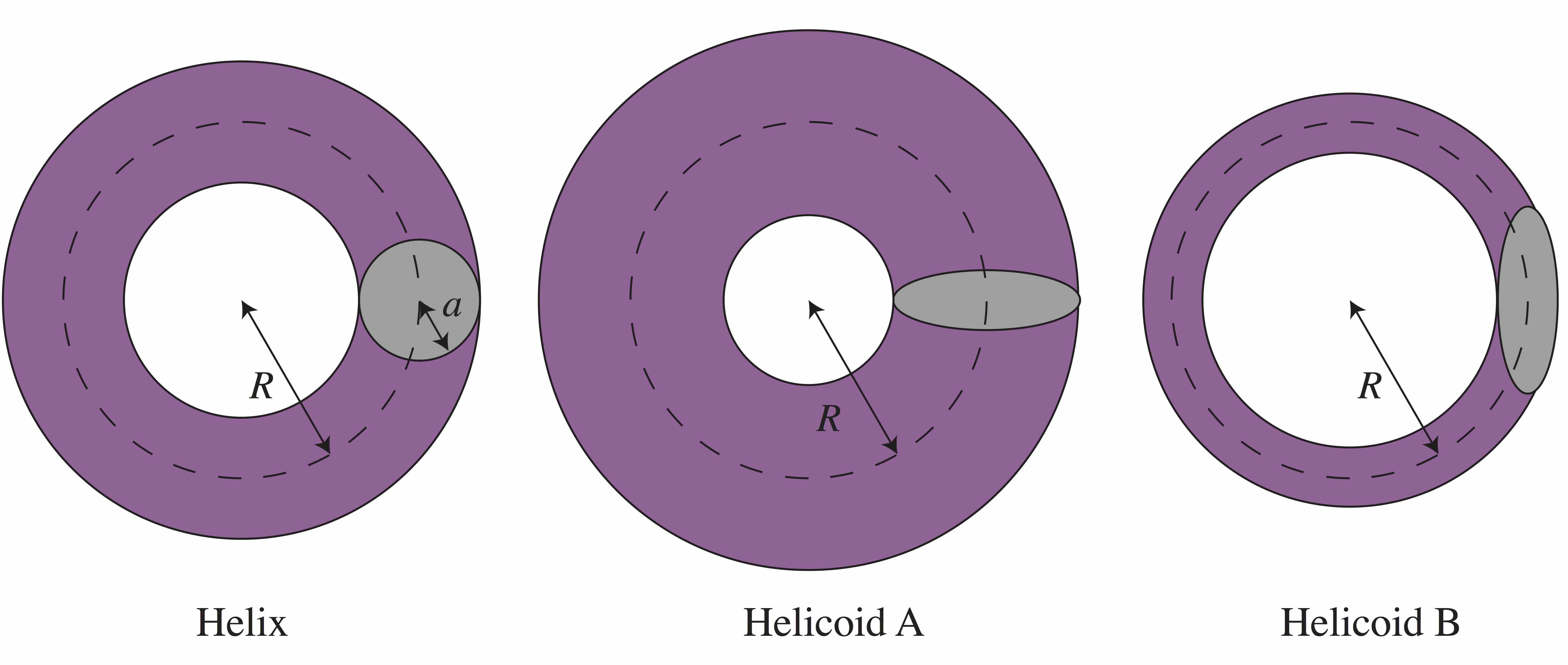}
\caption{\textbf{Cross-sections of the helical particles.}  }
\label{cross-section}
\end{figure}

We use physically extended helical nanoparticles. They are represented as composite particles with surface nodes that are treated as MD particles. The nodes are equally distributed on the surface with the center line following the parametric equation $(R \cos t, R \sin t, t L/(2\pi))$. The cross-sectional shapes are chosen to cover the extreme ranges of the flat shaped helices with elliptical cross-sections, helicoid A (B), that have the long-axis that are parallel (perpendicular) to the radial direction and the circular-cross section case, the helix Fig. \ref{cross-section}. The cross-sectional shape is given by $R/4(a_1\cos \phi,a_2\sin \phi)$ where $\phi\in[0,2\pi)$ and  $a_1=a_2=1$ for the helix, $a_1=1/2$ and $a_2=2$ for helicoids A and B. The node masses are set such that the total mass of the helix is equivalent to the mass of the displaced fluid and the area per node is given by 18.5 nm$^2$. 

\begin{figure}[htpb]
\centering
\includegraphics[width=0.45\textwidth]{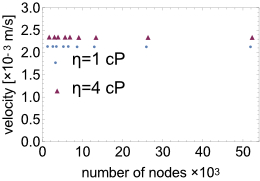}
\caption{\textbf{Node distribution dependence.} Propulsion velocities of helices with increasing number of nodes that at constant torque. 
}
\label{velocity node}
\end{figure}

In order to couple the MD particles to the fluid, the method introduces forces on both the MD particles and the fluid that result from an energy and momentum conserving interaction \cite{Ollila2011,mackay2013coupling}. These forces are calculated by assuming an elastic 3D collision between an MD node and a representative fluid particle. The representative fluid mass and momentum are interpolated at the location of the node that it interacts with using a Peskin interpolating stencil. 
\begin{eqnarray}
\vect{u}_f &=& \vect{u}_i + (\vect{v}_i-\vect{u}_i) \frac{2 m_v}{m_v+m_u},\\
\vect{v}_f &=& \vect{v}_i - (\vect{v}_i-\vect{u}_i) \frac{2 m_u}{m_v+m_u},
\end{eqnarray}
where $m_v$ and $m_u$ ($\vect{v}$ and $\vect{u}$) are the node and fluid mass (velocity), respectively \cite{mackay2013coupling}. An explicit no-slip boundary condition at the fluid-particle interface is implemented in all the simulations. This boundary condition sets the sum of the initial and the final relative fluid-node velocities to zero. All simulations are carried out with a lattice discretization of $\Delta x=13.245$ nm and $\Delta t=29.1859$ ps.

\subsection*{Implementation of the external torque}

The rotation of a helical particle in the simulations can be induced using two different methods. A constant torque is applied along the long axis of the helix. This corresponds to the synchronous regime wherein the frequency of rotation of the helix matches the frequency of rotation of the external magnetic field that exerts a torque on it \cite{morozov2014chiral}. In the absence of thermal fluctuations, 
this method is employed as a means to optimize the geometrical shape for the helical particle and 
the fluid viscosity that result in optimal propulsion and efficiency.


\subsubsection*{Acknowledgements}
This work was supported in part by the Academy of Finland through its Centres of Excellence Programme (2012-2017) under Project No. 251748 and Aalto Energy Efficiency Research Programme. We acknowledge the computational resources provided by Aalto Science-IT project and CSC-IT. The graphical representations in the Fig. \ref{directions} were rendered using VMD \cite{humphrey1996vmd}.


\end{document}